\documentclass[12pt]{spieman}  % 12pt font required by SPIE;
\usepackage{amsmath}
\usepackage{graphicx}
\usepackage{tocloft}
\usepackage{sidecap}
\newcommand\arcsec{\ensuremath{^{\prime\prime}}}

\cftpagenumbersoff{figure}
\cftpagenumbersoff{table}

\definecolor{ed}{rgb}{0.98, 0., 0.525}

\title{Measuring the Etalon Quality of the GREGOR Fabry-P\'erot Interferometer}

\author[*, 1]{Meetu Verma}
\author{Carsten Denker}
\affil{Leibniz-Institut f\"{u}r Astrophysik Potsdam (AIP), 
     An der Sternwarte~16,
     14482 Potsdam, 
     Germany}
     
\begin{document} 
\maketitle

%===============================================================================
%    ABSTRACT
%===============================================================================

\begin{abstract}
Imaging spectropolarimetry is an important observational tool in solar physics 
because of fast-cadence spectral scans with high-spectral resolution, large 
field-of-view, and its inherent suitability for post-facto image restoration. 
Fabry-P\'erot etalons are the key optical elements of these instruments. Their 
optical quality critically defines the instrument's performance. The two etalons 
of the GREGOR Fabry-P\'erot Interferometer (GFPI) were used for more than 
10~years, raising questions about the potential deterioration of etalons 
coatings. We present an assessment of the etalons optical quality, describe the 
inspection method based on Zernike polynomials, discuss the field dependence of 
the finesse and its consequences for instrument design, and investigate the 
impact of the measurement technique to achieve plate parallelism. We find that 
extended exposure to sunlight affects the etalon coatings, i.e., lowering the 
peak transmission and leaving an imprint of the pupil of the GREGOR solar 
telescope on the etalon that is directly exposed to sunlight. The finesse of 
both etalons, however, remains high so that the impact on imaging 
spectropolarimetry is negligible.
\end{abstract}

\keywords{Fabry-P\'erot, etalons, interferometers, spectrometers, optical 
inspection, coatings}    

{\noindent \footnotesize\textbf{*}Meetu Verma, \linkable{mverma@aip.de} }

\begin{spacing}{1}   % use double spacing for rest of manuscript

%===============================================================================
%    INTRODUCTION
%===============================================================================

\section{Introduction}
\label{INTRO} 

High-resolution solar physics mainly explores the visible and near-infrared 
spectral range to investigate flows and magnetic fields in the photosphere and 
chromosphere. Tracing changes on the solar surface across various atmospheric 
layers requires high spatial, temporal, and spectral resolution as well as good 
photometric and polarimetric accuracy. Currently, the largest solar telescopes 
have aperture diameters of 1.5\,--\,1.7~m. Photon-gathering capability and 
spatial resolution motivates pursuing telescopes with even larger apertures, 
i.e., the next generation of 4-meter aperture telescopes is already on the 
horizon. However, designing and building instruments for high-resolution solar 
physics is a challenging task because the science requirements are often 
irreconcilable, which leads to different instrument classes, e.g., classical 
spectrographs scanning the solar surface, spectrogrpahs employing integral 
field units (IFUs), and imaging spectrometers. Outfitted with polarimeters, all 
instruments can also perform measurements of the solar magnetic field.

Imaging spectropolarimetry is nowadays carried out with complex Fabry-Per\'ot 
interferometers (FPIs). Examples include the Triple Etalon SOlar Spectrometer 
(TESOS) \cite{Kentischer1998, Tritschler2002a} at the Vacuum Tower Telescope 
(VTT), the Interferometric BIdimensional Spectropolarimeter (IBIS) 
\cite{Cavallini2006} at the Dunn Solar Telescope (DST), the CRisp Imaging 
SpectroPolarimeter (CRISP) \cite{Scharmer2006a} at the Swedish Tower Telescope 
(SST), and the GREGOR Fabry-Per\'ot Interferometer (GFPI) \cite{Denker2010b, 
Puschmann2012} at the 1.5-meter GREGOR solar telescope \cite{Schmidt2012, 
Denker2012, Kneer2012} located at Observatorio del Teide, Tenerife, Spain. The 
next generation of FPIs is already on the horizon. The Visbile Tuneable Filter 
(VTF) \cite{Schmidt2014} offers a clear aperture with a diameter of about 
240~mm, a significant advancement in manufacturing large etalon plates, which 
are needed for the 4-meter Daniel K.\ Inouye Solar Telescope (DKIST) 
\cite{Tritschler2016} located at the Haleakal\={a} Observatory, Maui, Hawaii, 
U.S.A. The high photon-throughput of FPIs, especially when high spatial and 
spectral resolution are required at the same time, offers many advantages over 
classical spectrographs that have to scan the solar surface. In the latter case, 
this leads to slow cadences when scanning a region on the Sun at the 
diffraction-limited resolution of 4-meter aperture telescopes. More importantly, 
since FPIs deliver images, they benefit directly from real-time correction with 
adaptive optics (AO) systems \cite{Rimmele2006} and from post-facto image 
restoration techniques \cite{Loefdahl2007, Denker2015}.

The main optical design of the GFPI remained unchanged for more than a decade. 
However, outdated calibration equipment and advances in detector technology led 
to changes of the optical path for the laser alignment of the etalon plates. In 
addition, two synchronized CMOS cameras were recently integrated, which have a 
significantly higher image acquisition rate of up to 100~Hz. This motivated us 
to evaluate the quality and performance of the two Fabry-P\'erot etalons, which 
have been in operation in the GFPI for more than 10~years (delivery in March 
2005 and March 2007). A robust algorithm \cite{Denker2005a} was developed for 
measuring and maintaining the the plate parallelism of a Fabry-Per\'ot etalon 
used in the Visible-light Imaging Magnetograph (VIM) \cite{Denker2003a} at the 
Big Bear Solar Observatory (BBSO). Tuning the voltages that are applied to the 
piezoelectric actuators that control the tip-tilt motion of one etalon plate 
with respect to the other yields a direct relationship to the finesse of the 
etalon. The finesse across the plates was characterized using Zernike 
polynomials. An even more detailed investigation of the instrument profile was 
carried out for IBIS \cite{Reardon2008}. The current study presents similar 
tests and and expands the method to a dual-etalon FPI. Assessing the etalons has 
several goals: (1) demonstrate that the coatings are still of good quality after 
more than a decade, (2) characterize quantitatively the etalon performance, and 
(3) validate the current alignment strategy to ensure plate parallelism. 

In Section~\ref{GFPI}, we briefly describe the GFPI and current method to align 
the etalons and to measure the plate parallelism. Subsequently, we present in 
Section~\ref{EVAL} the setup in the VTT optical laboratory, where the etalons 
were relocated for assessment. Individual experiments and measurements will be 
outlined and reported in Section~\ref{RES}, which includes a description of the 
method, based on Zernike polynomials, which is used to characterize defects and 
quality of the etalon plates. We conclude our investigation in 
Section~\ref{DISC} with a discussion of the results and explain their impact for 
the continued operation of the GFPI at the GREGOR solar telescope.

%===============================================================================
%    GREGOR FABRY-PEROT INTEROMETER
%===============================================================================

\section{GREGOR Fabry-P\'erot Interometer}\label{GFPI} 

%---- general description of the GFPI ------------------------------------------
The GFPI is a tunable, dual-etalon system with a collimated mount, which is 
designed for high resolution imaging spectropolarimetry of the Sun. The etalon
coatings are optimized for the wavelength range 530\,--\,860~nm
\cite{Puschmann2012}. The spectral resolution depends on the observed
wavelength $\lambda$ and amounts to about ${\cal R} = \lambda / \Delta\lambda 
= 250\,000$, where $\Delta\lambda$ refers to the FWHM of the GFPI's 
transmission profile. Spectral lines are scanned sequentially with an image
acquisition rate of up to 100~Hz with a new camera system (10\,--\,20~Hz
until May 2019) and with a cadence of a few tens of seconds, while simultaneous
images in a broad-band channel facilitate image restoration of the 
spatio-spectral data cube. The new camera system has a higher duty cycle 
of about 90\%, i.e., the photometric accuracy will be significantly improved.
Fine-structures with sizes down to 60~km can be resolved in a 
field-of-view (FOV) of $63\arcsec \times 40\arcsec$ ($50\arcsec \times 
38\arcsec$ until May 2019) on the solar surface.

\begin{figure}
\center
\includegraphics[width=\textwidth]{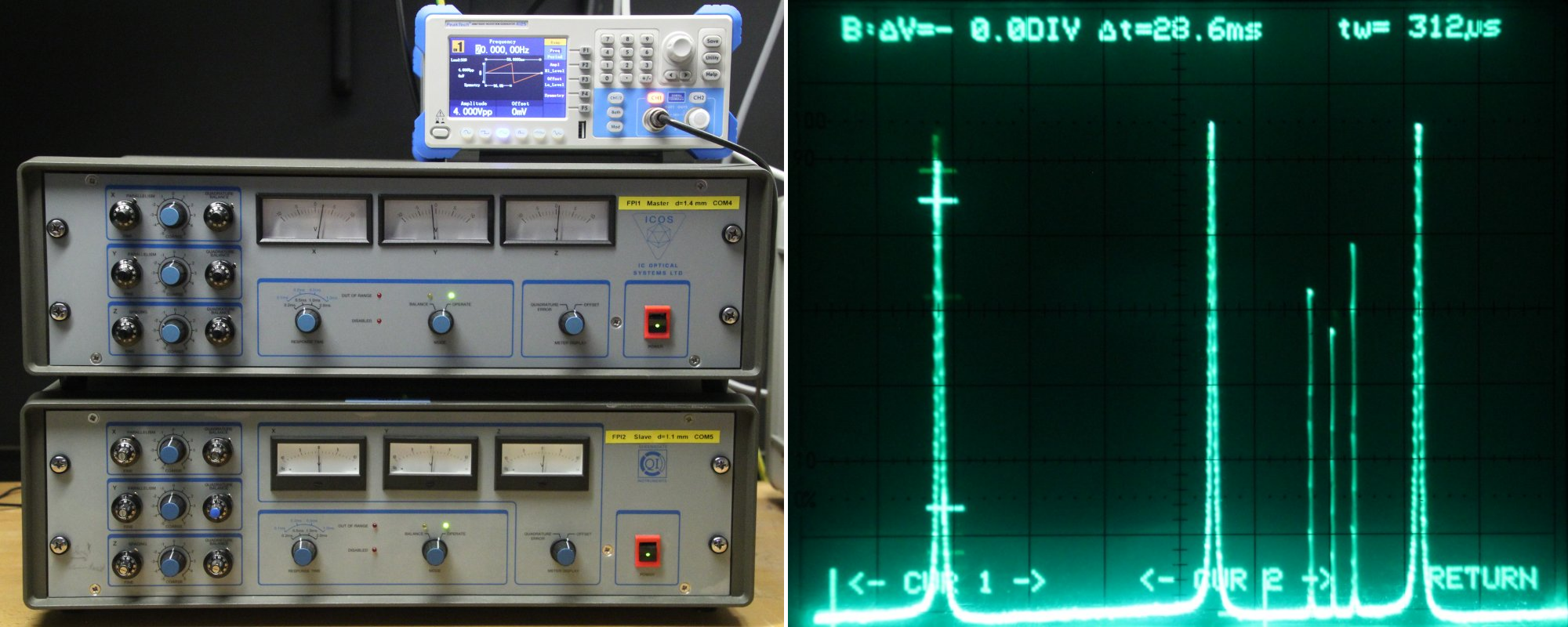}
\caption{Two ICOS CS100 controllers and the PeakTech 4125 waveform generator
    (\textit{left}). Screen of the Philips PM3350A digital storage oscilloscope
    (\textit{right}) displaying three transmission peaks, which were scanned
    with the ramp generator. The cluster of minor peaks results from
    resetting the scanning range. Finesse and FSR measurements are translated
    into a time measurement of the width (312~$\mu$s) and period of consecutive
    peaks (18~ms, not shown in the display), respectively, yielding an
    effective finesse of ${\cal F}_\mathrm{eff} = 57.7$ for FPI~2. The large
    cursors set the range ($\Delta t = 28.6$~ms) for determining the period,
    while the small cursors are used to derive a reference for the FWHM.} 
\label{FIG01}
\end{figure}

%---- etalons characteristics --------------------------------------------------
The GFPI etalons are in use since 2007 \cite{Puschmann2007}, first at the 
Vacuum Tower Telescope then starting in 2009 \cite{Denker2010b} at GREGOR. However, 
they were already purchased in 2005 (FPI~1, serial No.~1055) and 2007 (FPI~2,
serial No.~1057, first in the beam). Both etalons have narrow cavities with a 
plate spacing of $d_1 = 1.408$~mm and $d_2 =1.101$~mm, respectively. The main
characteristics of the two etalons are summarized in Table~\ref{TAB99}.
Maintaining the plate spacing and tuning are handled by two 12-bit CS-100
controllers (see left panel of Figure~\ref{FIG01}) manufactured by IC Optical
Systems (ICOS). In both etalons, capacitance sensors form a three-axis ($xyz$)
capacitance bridge, which ensures plate parallelism and maintains cavity 
spacing. The etalon plates can be tilted by changing the $x$- and $y$-settings
of the capacitance bridge. The etalons are tuned to a certain wavelength by
modifying the plate spacing via the $z$-setting. The conversion factors 
from a voltage to wavelength shift are $s_1 = 0.526$ and $s_2 = 0.681$~pm
mV$^{-1}$. The $xyz$-settings are either manually adjusted via coarse and
fine control knobs with millivolt precision or digitally via an RS-232 port,
which is used for scanning in wavelength. In addition, the output of a 
PeakTech 4125 waveform generator (left panel of Figure~\ref{FIG01} on top 
of the CS-100 controllers) can drive the $z$-axis, which is used in the 
alignment procedure of the etalon plates.

%-------------------------------------------------------------------------------
%    Table Etalon Characteristics
%-------------------------------------------------------------------------------
\begin{SCtable*}
\begin{tabular}{lcc}
\hline\hline
                                     &               FPI~1 & 
FPI~2\rule[-6pt]{0pt}{18pt}\\
\hline
Manufacturer                         &         ICOS \#1055 & ICOS 
\#1057\rule[0pt]{0pt}{12pt}\\
Date of measurements                 &        2018 July~25 & 2018 July~27\\
Diameter                             &               70~mm & 70~mm\\
Plate spacing $d$                    &            1.408~mm & 1.101~mm\\
Free spectral range (FSR)            &           0.1422~nm & 0.1819~nm\\
Reflectivity $R$ ($\lambda$632.8~nm) &             95.91\% & 94.99\%\\
Reflectivity finesse $F$             &               75.22 & 61.11\\
Conversion factor $s$                & 0.5265~pm mV$^{-1}$ & 0.6808~pm 
mV$^{-1}$\\
FWHM                                 &            1.890~pm & 2.976~pm\\
FWHM (measured)                      &            2.760~pm & 3.028~pm\\
Effective finesse $F_\mathrm{eff}$   &               51.52 & 
60.06\rule[-5pt]{0pt}{10pt}\\ \hline
\end{tabular}
\caption{Etalon characteristics. Note -- The numbering etalon numbering
    was switched compared to previous publications \cite{Puschmann2012}.
    However, this numbering scheme corresponds the current labels on the optics
    and controller, and it complies with the nomenclature of the GFPI control
    software.}
\label{TAB99}
\end{SCtable*}

%-------------------------------------------------------------------------------
%    Figure Controller & Figure Oscilloscope
%-------------------------------------------------------------------------------
%-------------------------------------------------------------------------------

The spacing of the etalons can be controlled via the RS-232 interface, which 
means that a certain voltage range is discretized at 12-bit resolution in 4096 
wavelength steps. FPI~1 serves as the master in the scanning process and the 
wavelength shift of FPI~2 is slaved to it so that their transmission peaks 
overlap when scanning solar spectral lines. Some older reference values for the 
wavelength step $\delta\lambda$ are listed in Table~\ref{TAB04}, where 
$\delta\lambda^\prime$ results from a linear fit to the measurements at four
wavelength positions. The corresponding linear model is given by
$\delta\lambda^\prime = c_0 + c_1 \cdot \lambda$ with $c_0 = 1.2800 \times
10^{14}$~m and $c_1 = 4.5807 \times 10^{-7}$, where $\lambda$ and
$\delta\lambda^\prime$ are given in meter (GFPI Technical Report No.~1, 2009
August~30). These values still apply to the current GFPI setup. Thus, at the
laser wavelength a single wavelength step of the GFPI corresponds to
$\delta\lambda^\prime = 0.3026$~pm, which significantly oversamples solar 
spectral lines.

\begin{SCtable*}
\begin{tabular}{lcc}
\hline\hline
Spectral Line & $\delta\lambda$ & $\delta\lambda^\prime$\rule[-6pt]{0pt}{18pt}\\
\hline
Fe\,\textsc{i} $\lambda$617.3~nm & 0.2954~pm & 0.2956~pm\rule[0pt]{0pt}{12pt}\\
Fe\,\textsc{i} $\lambda$630.3~nm & 0.3016~pm & 0.3015~pm\\
H$\alpha$ $\lambda$656.3~nm      & 0.3137~pm & 0.3134~pm\\
Fe\,\textsc{i} $\lambda$709.0~nm & 0.3375~pm & 0.3376~pm\rule[-5pt]{0pt}{10pt}\\
\hline
\end{tabular}
\caption{Wavelength step $\delta\lambda$ (measured) and $\delta\lambda^\prime$
    (interpolated) of the GFPI determined with the VTT echelle spectrograph on
    2009 April~11.}
\label{TAB04}
\end{SCtable*}

%---- laser channel and measuring procedure ------------------------------------
The etalons have to be exactly perpendicular to the incident collimated beam of 
sunlight. Typically once per observing season FPI~2 is removed from the beam and 
the retro-reflection from the front side of FPI~1 is traced back to the science 
focus F4 of the GREGOR telescope (auto-collimation). Stopping down the iris 
field stop in F4 creates a pinhole and perfect tip-tilt alignment is achieved 
once the retro-reflection from FPI~1 leaves the optical system via the pinhole. 
The procedure is repeated for FPI~2 once it is re-inserted into the beam and 
after blocking the optical train just behind FPI~2. This manual alignment of 
the etalons is accurate to about one minute of arc. Inter-etalon reflections 
are minimized by placing a high-quality neutral density filter with a 
transmission of about 70\% between the two etalons \cite{Bello2008}. A 
slight misalignment of the two etalons with respect to the optical axis leads to 
ghost images, which are manually removed  by minute tip-tilt corrections of 
FPI~1 while imaging a large pinhole. These adjustments are on the order of about 
10 seconds of arc.

\begin{figure}[ht]
\center
\includegraphics[width=\textwidth]{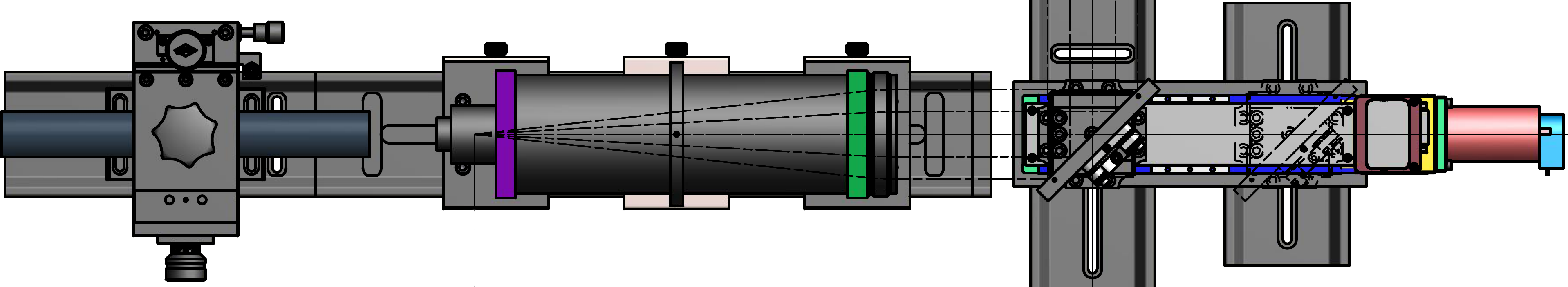}
\caption{The GFPI laser channel consists of (\textit{from left to right}) 
    a stabilized HeNe laser (632.8~nm), a beam expander consisting of a ball 
    lens (\textit{to left of the violet optics mount}) and collimator lens
    (\textit{to the right of the green optics mount}), and a motorized mirror,
    which is shown in the inserted position while the parking position is to 
    the far right. The turning mirror directs the collimated laser beam with 
    a diameter of 40~mm into the main optical train of the GFPI for aligning 
    the etalons. Another motorized turning mirror sends the beam to a transfer
    optics that focuses the laser light onto a photo-multiplier (\textit{not 
    shown in this design drawing}).}
\label{FIG02}
\end{figure}

%---- description of the laser channel ----------------------------------------
The regular etalon laser alignment procedure is carried out every one to two 
weeks, which corresponds to the typical length of an observing campaign at the 
GREGOR telescope. A red ($\lambda$632.8~nm) SL~03-Series stabilized HeNe laser 
manufactured by SIOS Me{\ss}technik GmbH is used for this purpose, which has a 
compact design and a rapid warm-up period (about 15~min). The laser can be 
operated in either frequency or amplitude stabilized mode. The frequency 
stabilized mode is the best choice for the alignment procedure. Motor controlled 
mirrors are inserted in the narrow-band channel just in front of the first 
etalon and behind the camera lens. The first mirror inserts a collimated laser 
beam with diameter of $D = 40$~mm that is created by the beam expander shown in 
Figure~\ref{FIG02}, which consists of a small ball lens just in front of the 
laser and a collimator lens ($f = 310$~mm and $D = 80$~mm). Even though 
desirable, a larger beam diameter cannot be created because of limited space on 
the optical tables and inside the GFPI housing. This shortcoming motivated 
partly the laboratory measurements in the following sections. Finally, the 
second mirror directs the now converging laser beam to a small relay optics that 
focuses the light upon a photo-multiplier.

%-------------------------------------------------------------------------------
%    Figure Laser
%-------------------------------------------------------------------------------
%-------------------------------------------------------------------------------

%---- measuring procedure for the finesse --------------------------------------
The signals from the waveform generator and the photo-multiplier provide the 
inputs for a Philips PM3350A digital storage oscilloscope, which can analyze 
arbitrary waveforms. A simple switch allows to toggle back and forth between 
FPI~1 and FPI~2, i.e., the position of the transmission peak of one etalon is 
kept constant while that of the other etalon is ramped across this position. The 
width of the laser line is certainly below 0.2~fm, and the manufacturer provides 
frequency values of 1\,--\,10~kHz for the line width, which indicate an even 
narrower laser line. Thus, a $\delta$-function is a very good assumption for the 
laser line in the following discussion. The beam expander creates a Gaussian 
beam, which illuminates first FPI~2. The plate separation of FPI~2 is changed 
via its $z$-voltage to maximize the peaks recorded with the digital 
oscilloscope. After that the $z$-position is kept constant. As a consequence, 
the Gaussian beam is modulated by the transmission function, and the blue-shift 
of FPI~2 introduces an additional variation of the intensity. This resulting 
illumination of FPI~1 is no longer symmetric and the coating of FPI~2 adds 
significant granularity to the intensity distribution. Ramping the $z$-voltage 
of FPI~1 exposes its transmission profile locally to varying intensities. As 
FPI~1 also exhibits a blue-shift in the collimated beam, local transmission 
profiles across the beam are shifted in wavelength. The recorded transmission 
profiles at the photo-multiplier averages all these contributions. Therefore, 
the measured effective finesse ${\cal F}_\mathrm{eff}$ (the calligraphic font 
indicates a measurement with the photo-multiplier) will be significantly lower 
compared to the locally derived finesse values across the etalon plates. If the 
etalons are properly aligned, then the display of the oscilloscope will show 
narrow and high transmission peaks (see right panel of Figure~\ref{FIG01}). This 
goal is accomplished by tuning the $xy$-settings of the CS-100 controllers for 
FPI~1, while keeping those for FPI~2 fixed, and vice versa. Taking turns in 
adjusting the $x$- and $y$-settings, the transmission peaks are maximized. Low 
and broad transmission peaks indicate that the etalon plates are still not 
parallel. In addition, the width of the transmission peak is continuously 
displayed in the digital storage mode so that a quantitative feedback is given 
as well. A quantitative description of this measuring procedure is given in 
Section~\ref{RES}. Using the digital memory of the oscilloscope, the width 
(effective finesse) and separation (free spectral range, FSR) of the 
transmission peaks are measured, which allows us to derive a measure of the 
effective finesse ${\cal F}_\mathrm{eff}$ in real-time. A snapshot of a typical 
measurement is shown in the right panel of Figure~\ref{FIG01}. Note that the 
separation of the peaks is determined in a different menu. The basic concept of 
the overall alignment procedure is to translate the quantitative assessment of 
the etalons' finesse into a time measurement of the two convolved transmission 
functions.

%==============================================================================
%    OPTICAL SETUP FOR THE ETALON EVALUATION
%==============================================================================

\section{Optical Setup for the Etalon Evaluation}
\label{EVAL}

%---- general remarks about the optical setup ---------------------------------
The characteristic evaluation of the etalons was carried out in one of the 
optical laboratories of the VTT. The laboratory is air conditioned 
and provides a stable environment for the experiments. The optical setup was 
placed on a large optical table using L95 optical rails for proper alignment of the 
mechano-optical components. 

\begin{figure}[ht]
\includegraphics[width=\textwidth]{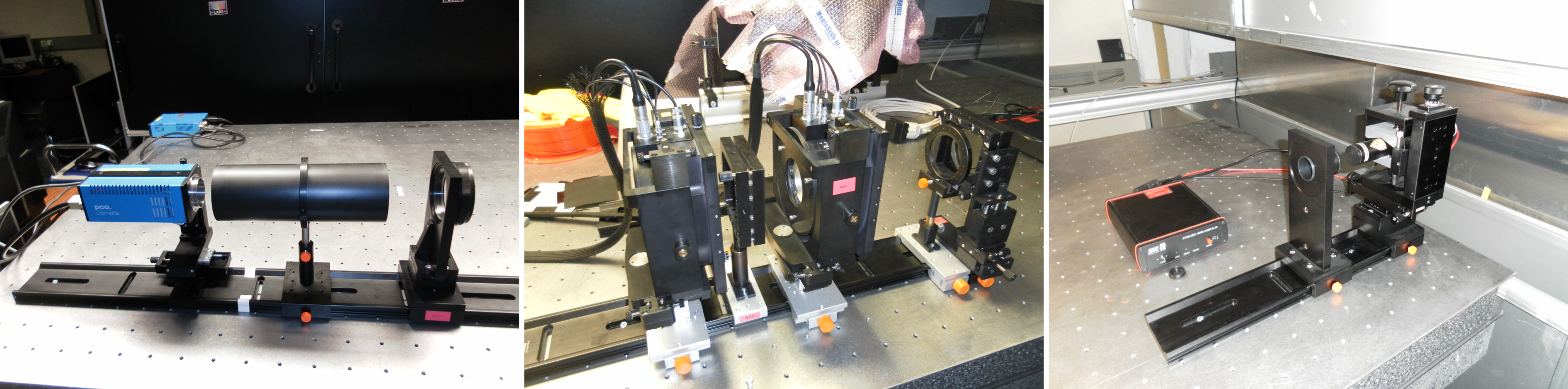}
\caption{The optical setup to measure etalon quality consists the laser beam
    expander (\textit{right}) followed by the collimator lens L1, iris stop,
    and the etalons with inserted high-quality neutral density filter
    (\textit{middle}), and finally the camera lens L2 and pco.4000 CCD
    camera (\textit{left}).}
\label{FIG03}
\end{figure}

%---- optical setup -----------------------------------------------------------
The frequency-stabilized HeNe laser, five-axis laser mount, and ball lens were 
taken from the GFPI setup at the GREGOR telescope. The optical setup 
(Figure~\ref{FIG03}) produced a collimated beam for measuring the surface 
quality of the etalons. The achromatic lens L1 with a focal length of $f_1 = 800$~mm 
yielded a much larger beam than in the laser channel of the GFPI, and was 
stopped down to a diameter of $D = 65$~mm by a large-aperture iris diaphragm. 
This diameter was chosen slightly smaller than the free aperture of the etalons 
(70~mm) to avoid edge effects at the periphery of the etalon plates. Note 
that the focal length of this collimator lens differs significantly from that 
of the GFPI laser channel. A test target, which consisted of several USAF 1951 
resolving power test targets in combination with a millimeter grid, established 
the exact position for examining the etalons. Another achromatic lens L2 with a 
focal length of $f_2 = 310$~mm (i.e., the original collimator lens of the GFPI 
laser channel) was inserted to image the test target on the CCD detector.

%-------------------------------------------------------------------------------
%    Figure optical setup
%-------------------------------------------------------------------------------
%-------------------------------------------------------------------------------

\begin{figure}[ht]
\centering
\includegraphics[width=0.95\textwidth]{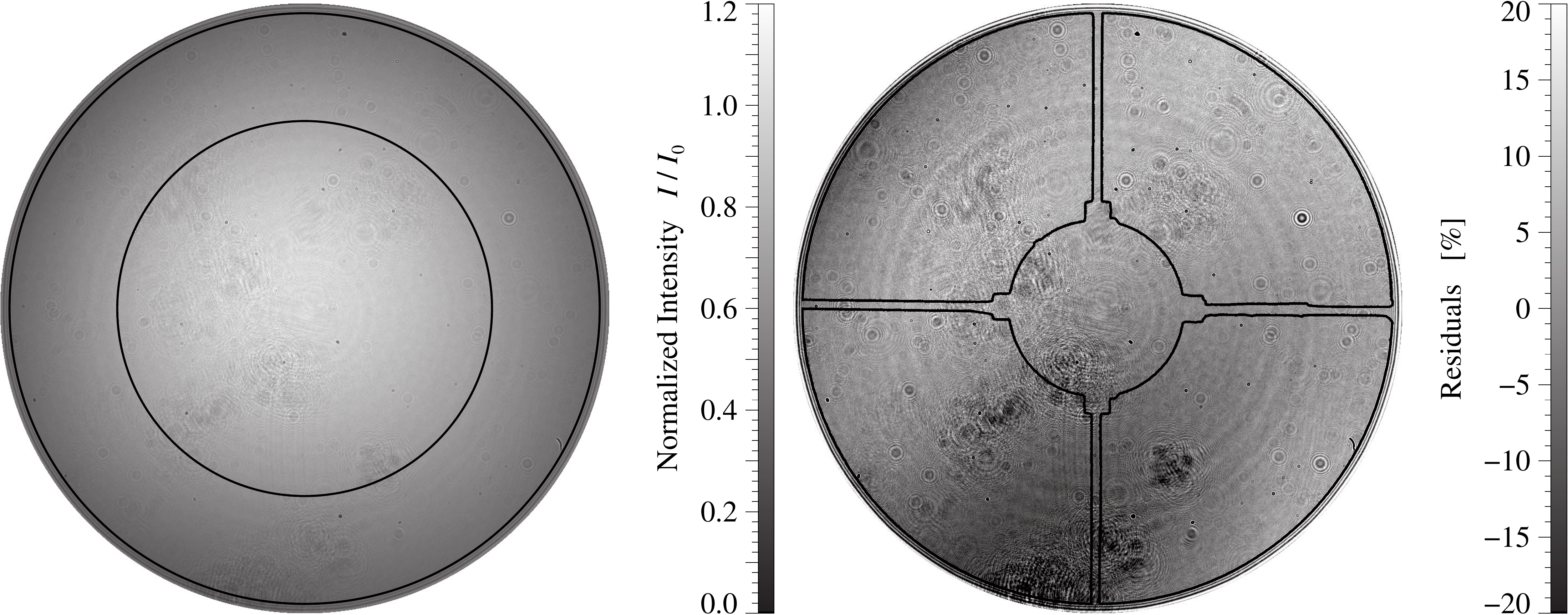}
\caption{Collimated Gaussian laser beam with interference fringes and 
    artifacts from dust particles on the entrance window of the CCD camera
    (\textit{left}) and the corresponding residuals after subtracting a
    two-dimensional Gaussian (\textit{right}). The beam diameter is 
    $D = 65$~mm, and the two solid circles in the left panel refer 
    to diameters of 63~mm (pupil) and 40~mm (GFPI laser beam). The entrance 
    pupil of the 1.5-meter GREGOR solar telescope was superimposed on the
    residuals in the right panel, where the outer diameter corresponds 
    to 63~mm and that of the central obscuration to 18.5~mm.}
\label{FIG04}
\end{figure}

%---- CCD detector -------------------------------------------------------------
The experimental data were recorded with a pco.4000 CCD camera -- one of the 
facility cameras at the VTT. This large-format detector has a resolution of 4008 
$\times$ 2672 pixels, a pixel size of 9~$\mu$m $\times$ 9~$\mu$m, and linear 
dimensions of 36~mm $\times$ 24~mm. A smaller region-of-interest (ROI) with 2560 
$\times$ 2560 pixels was extracted from the full-format frames, which 
encompasses the Gaussian intensity distribution of the laser beam. The diameter 
of the beam covers 2276~pixels on the detector, which results with $D = 65$~mm 
in an image scale of 28.6~$\mu$m pixel$^{-1}$. The left panel of 
Figure~\ref{FIG04} shows the Gaussian laser beam. The superimposed outer circle 
refers to the pupil size in the GFPI narrow-band channel, whereas the inner 
circle circumscribes the beam in the laser channel of the GFPI, which is used to 
ensure the parallelism of the etalon plates. A two-dimensional Gaussian with a 
FWHM = 52~mm was fitted to the beam and the residuals after subtraction from the 
beam are depicted in the right panel of Figure~\ref{FIG04}. Artifacts are 
clearly discernible in this depiction, arising mostly from the cover window of 
the CCD camera. The contour lines represent the geometry of the entrance pupil 
of the GREGOR telescope, outlining the support structure of the secondary mirror 
and the obscuration caused by the secondary mirror. The exact orientation of the 
pupil image is time dependent because the alt-azimuth mount of the GREGOR 
telescope introduces image and pupil rotation. Even though the experiments 
were carried out in the optical laboratory of the VTT, the GREGOR pupil is 
presented for reference because ultimately the characterization of the etalons 
must be interpreted in the context of GFPI operations at the GREGOR telescope.

%-------------------------------------------------------------------------------
%    Gaussian Laser Beam
%-------------------------------------------------------------------------------
%-------------------------------------------------------------------------------

The pco.4000 cameras have two A/D converters for faster readout and higher 
image acquisition rates. However, since high-cadence imaging was not required, only a 
single A/D converter was selected to avoid an imbalance of the gain between the 
two halves of the detector. Image frames were recorded at a rate of 2.8~Hz 
without binning and with a dynamic range of 14~bit. 

%---- calibration data ---------------------------------------------------------
Dark frames were acquired by simply covering the CCD detector with a screw-on 
metal cap. In general, the light level in the optical laboratory was so low 
that there is no noticeable difference between the dark counts of the covered and 
uncovered detector. Typically, 20 dark frames are averaged to minimize the 
thermal and statistical variations inherent to the dark signal. Before 
inserting one or both etalon(s) into the collimated laser beam, 100 images of the 
unobscured Gaussian beam were recorded, which serve as a pseudo flat-field 
frame for the measurements with inserted the etalon(s).

%-------------------------------------------------------------------------------
%    RESULTS
%-------------------------------------------------------------------------------

\section{Results}
\label{RES}

\begin{figure}[ht]
\centering
\includegraphics[width=0.95\textwidth]{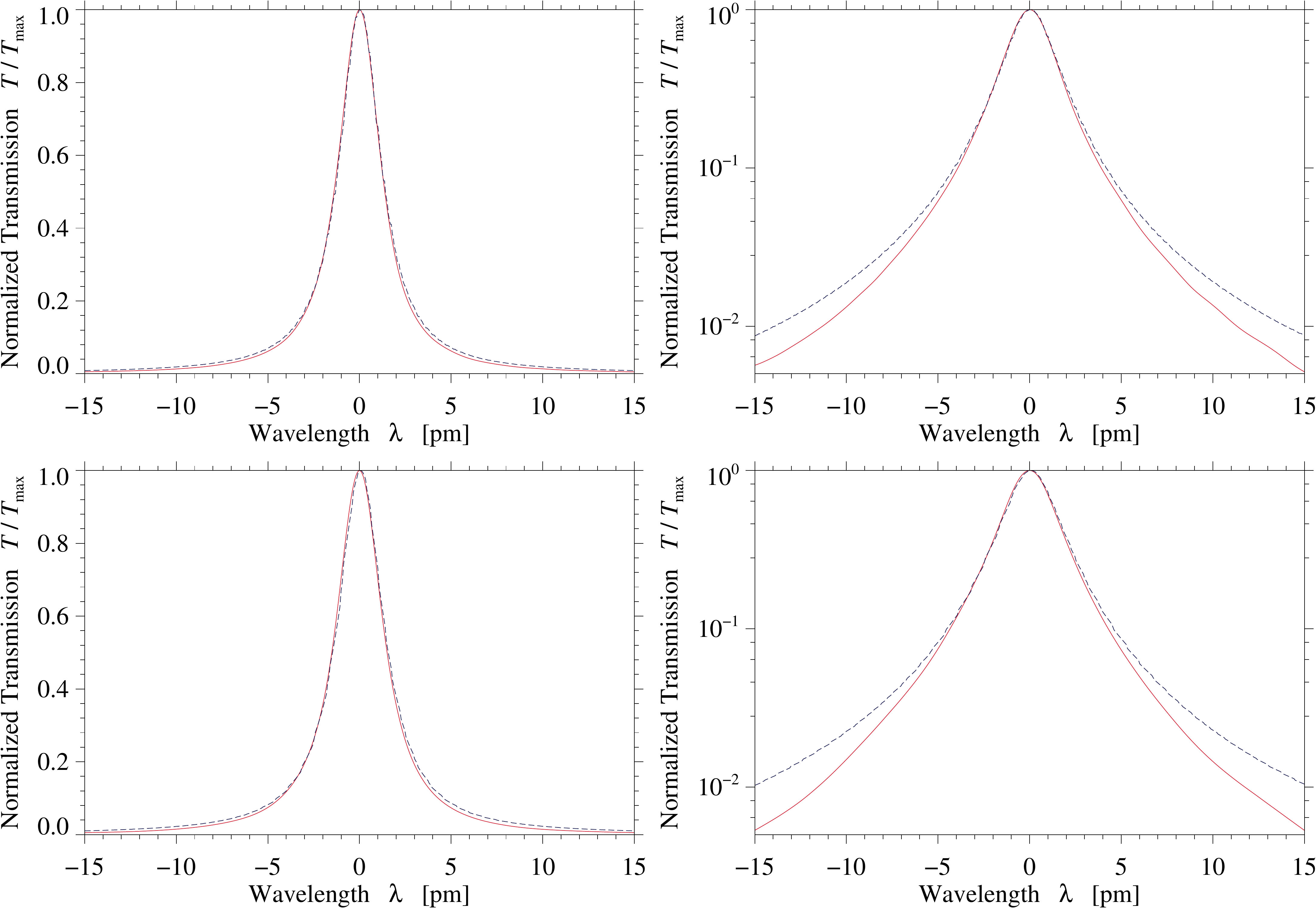}
\caption{Transmission profiles of FPI~1 (\textit{top}) and FPI~2
     (\textit{bottom}) plotted on a linear (\textit{left}) and a logarithmic 
     (\textit{right}) scale. The observed transmission profiles (\textit{solid red})
     are characterized by slightly narrower FWHM and significantly lower
     transmission in the wings as compared to fits with analytic Airy
     functions (\textit{dashed blue}).}
\label{FIG05}
\end{figure}

% !!!Transmission curves!!!
A thorough introduction to Fabry-P\'erot etalons is given in the classical 
monograph \textit{Principles of Optics} \cite{Born1998} and more recently in a 
detailed study \cite{Bailen2019}, which compared the collimated and telecentric 
mount options for etalons with an emphasis on their point spread functions 
(PSFs) and their performance in measuring Doppler velocities and magnetic 
fields. The transmission curve of a Fabry-P\'erot etalon is given by the Airy 
function
\begin{equation}
I = \left(1-\frac{A}{1-R}\right)^2 \left (1+F \sin^2 \delta/2 \right)^{-1}\ 
\end{equation}
with 
\begin{equation*}
F = \frac{4R}{(1-R)^2} \quad \mathrm{and} \quad
\delta = \frac{4 \pi n d \cos \theta}{\lambda},
\end{equation*}
where $I$ is the transmitted intensity, $A$ indicates absorption of the 
coating, $R$ represents the reflectivity of the coating, $F$ is the finesse coefficient, 
$\delta$ is the phase difference, $n$ is the index of refraction, $d$ is the 
plate spacing of the etalon, $\theta$ is the angle of incidence, and $\lambda$ 
is the observed wavelength. The average transmission profile of the etalons
is calculated from the local transmission profiles for each pixel contained
within the 65-mm aperture after removing the blue-shift, i.e., by aligning
the individual transmission peaks. This is accomplished by centering the local
profiles using linear interpolation before taking the average. The measured
transmission curve of both FPIs is fitted with an Airy function. The results 
are displayed in Figure~\ref{FIG05}. The wings of the Airy functions exhibit 
a higher transmission than the measured curves. The effective finesse for 
FPI~1 and~2 is $F_{\mathrm{eff},1} = 51.5$ and $F_{\mathrm{eff},2} = 60.6$,
respectively, which exceeds the conservative value of $F_{\mathrm{eff}} = 46$
provided by the manufacturer of the etalons. However, the effective finesse
$F_\mathrm{eff}$ of FPI~1 is lower than that of FPI~2 (see Table~\ref{TAB99}),
which disagrees with calculations based on the coating curves provided by the
manufacturer. This discrepancy is also confirmed by the manual finesse
measurements during the GFPI laser alignment procedure at the GREGOR telescope.
In the later case, the smearing of the average transmission profile because of
the blue-shift will lower the finesse values, while limiting the measurements
to the central part will raise them. While the measurement differ in detail 
for both cases, they are in general compatible. Using the logarithm of the
transmission curve as input for the fit will result in a better match for
the wings. However, the central transmission peak shows stronger deviations
in this case, which is not acceptable. Using different weighting functions 
will also yield slightly different finesse values. Thus, an error of a few 
percent has to be expected for determining the finesse.

%-------------------------------------------------------------------------------
%    Figure Etalon Transmission profiles
%-------------------------------------------------------------------------------
%-------------------------------------------------------------------------------

% blue shift and fitting procedure for the transmission peaks
The wavelength shift across the collimated laser beam, i.e., the so-called 
blue-shift, is used as a measure of the plate parallelism. The observed 
wavelength shift is derived by fitting the transmission profiles at each pixel 
with a Lorentzian \cite{Markwardt2009}, which yields peak transmision, 
position, and FWHM of each transmission peak. The Lorentzian is very good
approximations of the Airy function in proximity to the transmission peak
\cite{Sloggett1984}. Note that Gaussian fits to the measured 
transmission profiles are inaccurate, i.e., they underestimate the peak 
transmission by about 10\% and overestimate the FWHM by more than 25\%. This 
should be taken into account when comparing finesse measurements reported in 
literature. Thus, Gaussian fits significantly undervalue the quality of 
Fabry-P\'erot etalons. The averages of the characteristic values are computed 
for the full beam with a diameter of 65~mm, for the reduced beam with a 
diameter of 40~mm, and for the pupil mask of the GREGOR telescope with a 
diameter of 63~mm, which includes the obscuration of the secondary mirror and 
its support structure. The diameter of 40~mm corresponds the beam size in the 
GFPI laser channel, which had to be compact due to the limited space inside
the GFPI housing. The 63-mm GREGOR pupil is the typical beam shape during GFPI 
science observations, and the 65-mm free aperture provides a reference for 
observations with an unobstructed telescope. The motivation for showing all 
three cases is to validate the alignment procedure of the etalon plates 
irrespective of the shape and size of beam.

%-------------------------------------------------------------------------------
%    Figure 
%-------------------------------------------------------------------------------
\begin{figure}[th]
\centering
\includegraphics[width=0.95\textwidth]{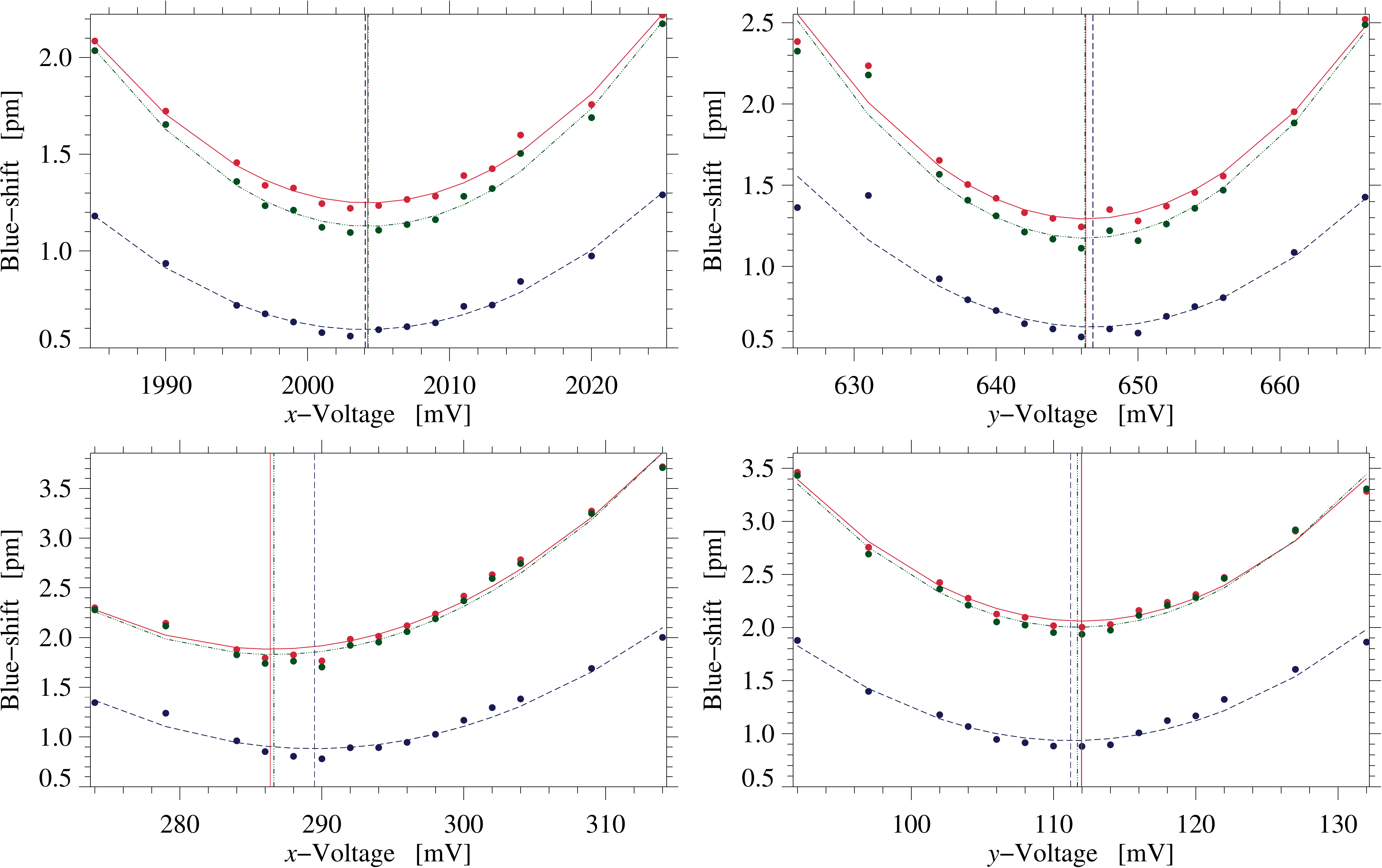}
\caption{Average blue-shift measured across FPI~1 (\textit{top}) and FPI~2
    (\textit{bottom}) as a function of the voltage applied to the
    piezoelectric actuators for the $x$- and $y$-direction. Blue-shifts were
    computed for the full 65-mm beam (\textit{solid red}), the 40-mm GFPI laser 
    channel beam (\textit{dashed blue}), and the 63-mm GREGOR pupil mask
    (\textit{dash-dotted green}). The colored dots refer to measurements, and the curves represent second-order polynomial fits. The minimum values for all three cases are indicated by vertical lines.}
\label{FIG06}
\end{figure}. 
%-------------------------------------------------------------------------------

% alignment procedure
The blue-shift measurements are carried out by manually changing the voltages of 
the piezoelectric actuators controlling the $x$- and $y$-parallelism of the 
etalon plates. The scale of the voltage potentiometers is given in steps of 
2~mV, i.e., manual adjustments can only achieve a precision of 1~mV. Initially, 
a voltage range of 40~mV around the transmission peak is sampled with 
non-uniform voltage increments, where the finest sampling of 2~mV covers the 
core of the transmission peak. The rough position of the peak can be gauged 
visually so that the independent scans for the $x$- and $y$-direction are 
typically already well centered with respect to the orthogonal direction. This 
results in 15 scans of the transmission curve for the $x$- and $y$-directions. 
In principle, a two-dimensional blue-shift map can be determined for each of the 
30 scans. A second-degree polynomial is fitted to the data for the $x$- and 
$y$-voltage (Figure~\ref{FIG06}). The parabolic curves for the $x$- and 
$y$-directions are slightly different for both FPIs, which could be an 
indication of minor differences in the response of piezoelectric actuators. The 
blue-shift curves for the full beam and that for the pupil mask are almost 
identical. In contrast, the blue-shift values for the reduced 40-mm beam are 
much lower because it only samples the central part of the etalon plates. The 
minimum of the blue-shift and the corresponding voltage are saved to perform the 
final transmission curve scans with a finer resolution of 1~mV in the applied 
voltage. For both FPIs and in all three cases, the $x$- and $y$-voltages 
corresponding to the minimum blue-shift are virtually identical 
(Table~\ref{TAB02}). The only exception is the $x$-voltage for FPI~2, which can 
be attributed to the sub-optimal centering of the parabola. In addition, the 
manual adjustment of the voltages can introduce errors, which easily explain the 
observed deviations from the parabolic fits. However, taking these limitations 
into consideration, the presented alignment procedure \cite{Denker2005a} is very 
robust and leads to almost identical results for the optimal voltage settings 
that ensure the parallelism of the etalon plates.

%-------------------------------------------------------------------------------
%   Table 2 
%-------------------------------------------------------------------------------
\begin{table}[ht]
\centering
\caption{Blue-shift parameters and voltages for both FPIs and for three beam 
    characteristics.}
\footnotesize \smallskip  
\begin{tabular}{lccccc}
\hline\hline
 & \multicolumn{5}{c}{\textbf{FPI~1}} \rule[-6pt]{0pt}{18pt}\\
\hline
& \multicolumn{2}{c}{$x$-direction} & & 
\multicolumn{2}{c}{$y$-direction}\rule[-5pt]{0pt}{16pt} \\
\cline{2-3} \cline{5-6}
   & Voltage [mV] & Shift [pm] & & Voltage [mV] & Shift 
[pm]\rule[-5pt]{0pt}{16pt} \\
\hline             
65~mm        & 2004.2  & 1.25  & & 646.3 & 1.29 \rule[0pt]{0pt}{11pt}\\
40~mm        & 2004.1  & 0.60  & & 646.8 & 0.63\\
GREGOR pupil & 2004.3  & 1.13  & & 646.3 & 1.17\rule[-5pt]{0pt}{10pt}\\
\hline
 & \multicolumn{5}{c}{\textbf{FPI~2}} \rule[-6pt]{0pt}{18pt} \\
\hline
& \multicolumn{2}{c}{$x$-direction} & & 
\multicolumn{2}{c}{$y$-direction}\rule[-5pt]{0pt}{16pt} \\
\cline{2-3} \cline{5-6}
   & Voltage [mV] & Shift [pm] & & Voltage [mV] & Shift 
[pm]\rule[-5pt]{0pt}{16pt} \\
\hline             
65~mm        & 286.3  & 1.88  & & 112.0 & 2.06\rule[0pt]{0pt}{11pt}\\
40~mm        & 289.5  & 0.88  & & 111.2 & 0.94\\
GREGOR pupil & 286.6  & 1.83  & & 111.7 & 2.00\rule[-5pt]{0pt}{10pt}\\
\hline
\end{tabular}
\label{TAB02}
\end{table}
%-------------------------------------------------------------------------------

\begin{figure}[ht]
\center
\includegraphics[width=\textwidth]{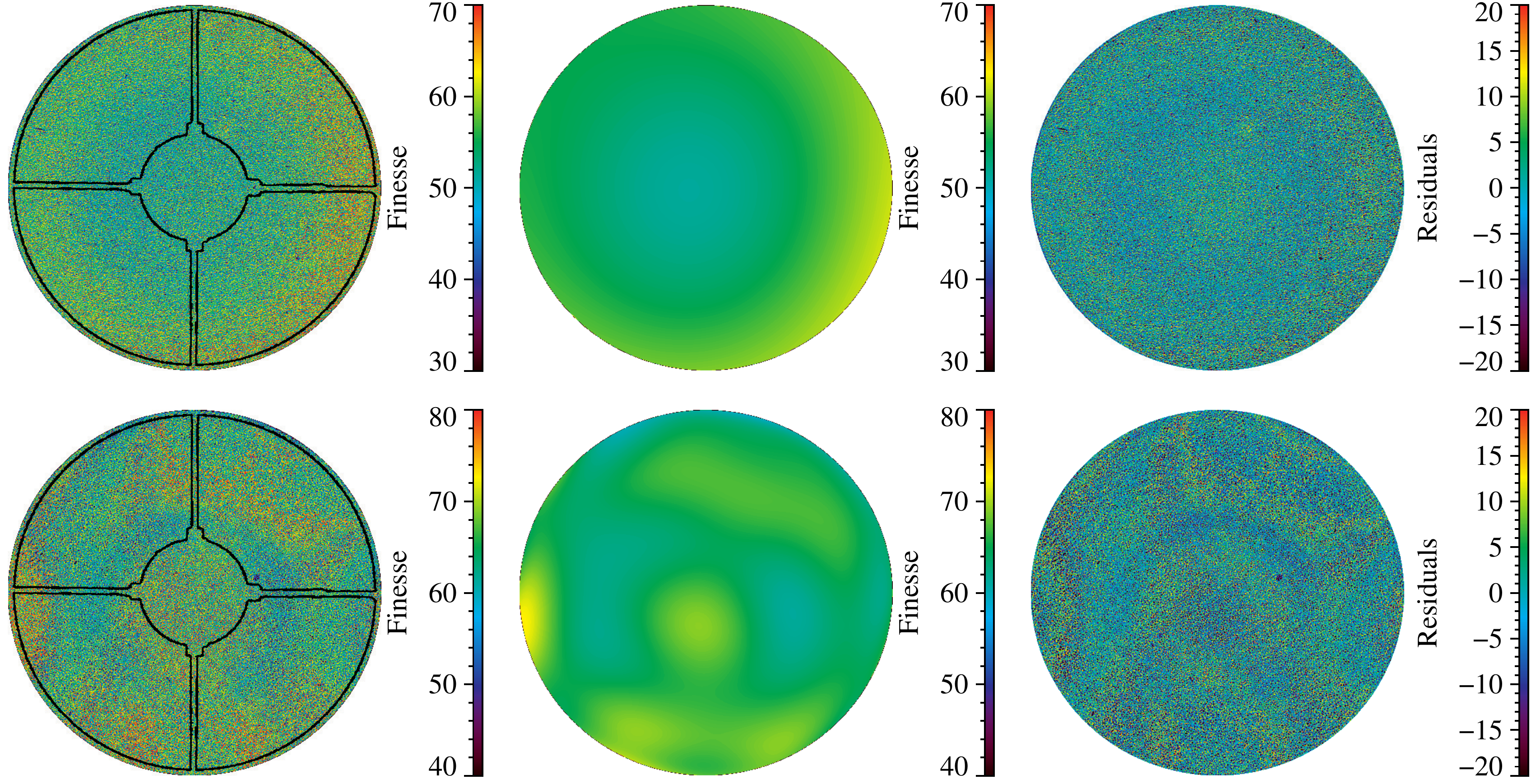}
\caption{Effective finesse across the FOV of FPI~1 (\textit{top}) and FPI~2
    (\textit{bottom}). The original measurements (\textit{left}), the
    corresponding Zernike polynomial fit (\textit{middle}), and the residuals
    of the fit (\textit{right}). Note that the finesse range is not the same
    for FPI~1 and FPI2. The entrance pupil of the 1.5-meter GREGOR solar 
    telescope was superimposed on the orginal finesse measurements 
    (\textit{left panels}).}
\label{FIG07}
\end{figure}

Once the etalon plates are aligned using the values of Table~\ref{TAB02}, the 
transmission profiles are scanned with 64 equidistant steps in the $z$-direction 
and a resolution of 1~mV. The blue-shifts are again computed from Lorentzian 
fits to the transmission profiles at each pixel. The measured finesse for both 
FPIs is shown in the left panels of Figure~\ref{FIG07}, whereas the middle 
panels depict a representation of the finesse in Zernike polynomials, and the 
right panels display the residuals after subtraction of the two-dimensional fit 
from the measured data. Zernike polynomials were already previously used to 
describe the finesse and other characteristic parameters of Fabry-P\'erot 
etalons \cite{Denker2005a}. This set of orthogonal polynomials was introduced by 
the optical physicist Frits Zernike \cite{Zernike1934} to represent the 
expansion of a wavefront for optical systems with circular pupils. The 
polynomials are the product of angular functions and radial polynomials defined 
on a unit circle. In this work, the Noll-ordering scheme \cite{Noll1976} is 
used, where the mode-ordering number $j$ is a function of radial degree $n$ and 
azimuthal frequency $m$. In the following, the two-dimensional maps of the 
characteristics parameters were constructed from Zernike polynomials with a 
mode-ordering number up to $j=55$, which corresponds to a radial degree $n \leq 
10$ and azimuthal frequency $|m| \leq 10$. 

%-------------------------------------------------------------------------------
%    Figure finesse FPI~1 and FPI~2
%-------------------------------------------------------------------------------
%-------------------------------------------------------------------------------

\begin{SCfigure*}
\includegraphics[width=0.5\textwidth]{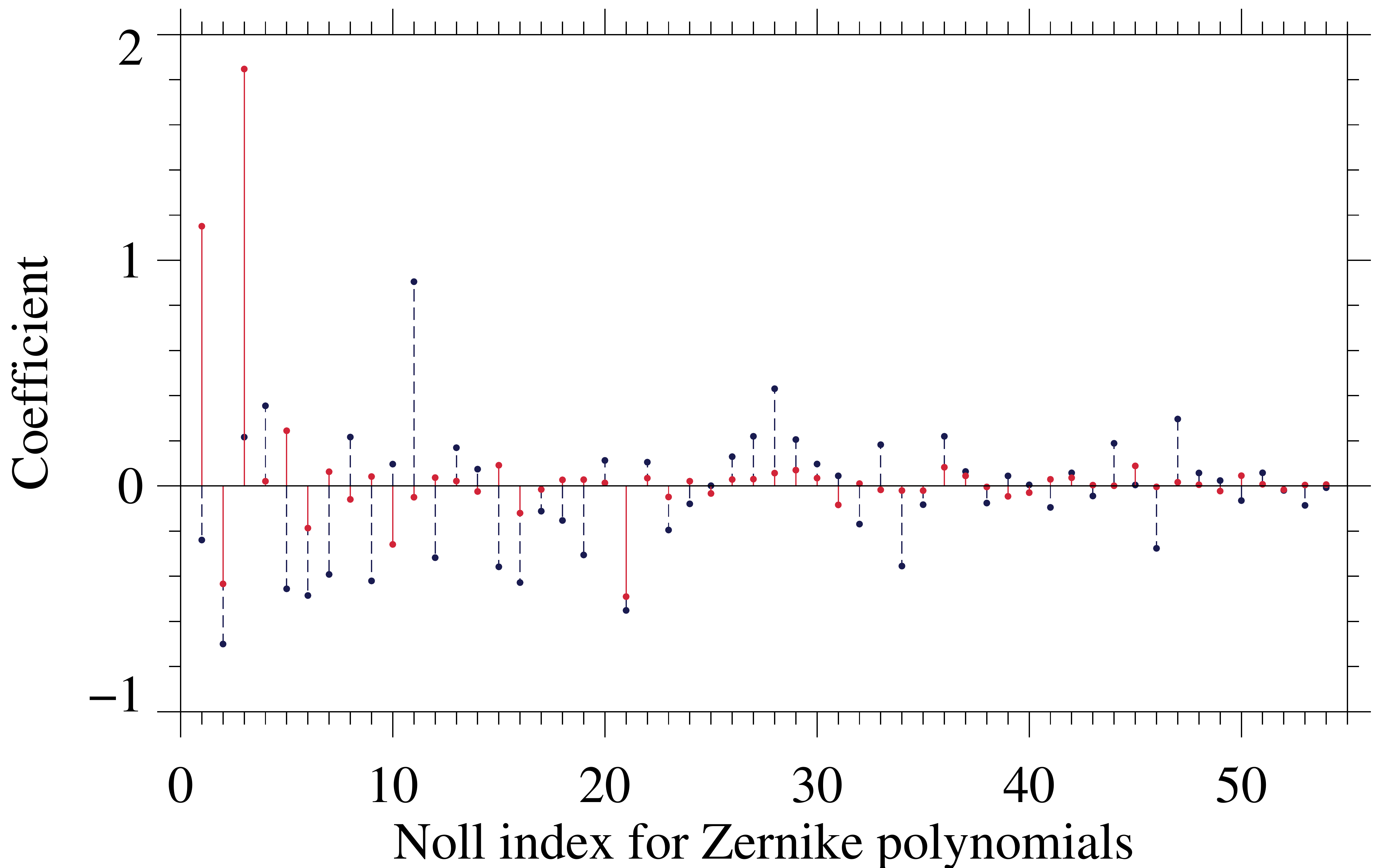}
\caption{Coefficients of 55 Zernike polynomials fitted to the finesse 
of FPI~1 (\textit{solid red}) and FPI~2 (\textit{dashed blue}), 
respectively.}
\label{FIG08}
\end{SCfigure*}

The two-dimensional maps of the finesse of both FPIs (left panels of 
Figure~\ref{FIG07}) are fitted with Zernike polynomials up to mode-ordering 
number $j=55$. The fitted coefficients are displayed in Figure~\ref{FIG08} as 
red and blue bullets with vertical bars to enhance legibility. The coefficients 
of FPI~1 become smaller with increasing $j$ so that the fit can be terminated at 
$j=15$. In contrast, the coefficients of FPI~2 remain significant up to $j=47$ 
suggesting a limit of $j=55$. The first coefficient belongs to the piston term, 
which is essentially the mean value of the effective finesse $F_\mathrm{eff} = 
55.1$ and 65.2 for FPI~1 and FPI~2, respectively. The coefficients of FPI~1 
representing tilt, tip, and oblique astigmatism are significant, i.e., the 
surface quality of the etalon is characterized by low-order deviations from 
perfectly parallel plates. In contrast, the coefficients of FPI~2 corresponding 
to coma, trefoil, and spherical astigmatism are large. 

Two-dimensional maps of the effective finesse are derived from the fitted 
Zernike coefficients and are displayed in the middle panels of in 
Figure~\ref{FIG07} along with the residuals in the right panels. Both FPIs have 
an atypical property, i.e., the higher finesse values are located at the outer 
edges and not in the center as expected from previously published measurements 
of etalon plates \cite{Denker2005a,Reardon2008,Bailen2019,Greco2019}. The 
finesse of FPI~1 has a symmetrical structure with low values in the central part 
and high values at the outer edges, whereas finesse of FPI~2 has high values in 
the central part matching the central obscuration of the GREGOR pupil 
surrounded by an annulus of lower finesse values and accompanied by 
significant variations of the finesse at the periphery of the etalon plates. The 
ring-like structure and variation along the perimeter could be caused by the 
non-uniformly rotating GREGOR pupil. However, original fabrication errors or 
plates distortions over time cannot be ruled out. The complex pattern of the 
finesse observed for FPI~2 justifies the choice of a higher limit $j=55$ for 
Zernike polynomials. The residuals of the fit show a granular pattern for both 
FPIs indicative of the micro-roughness and the coating inhomogeneities of the 
etalon plates. The measured effective finesse values for the three beam settings 
are compiled in Table~\ref{TAB03}. The mean values of the effective finesse 
value are similar for all three beams, e.g., choosing the GREGOR pupil yields 
$F_\mathrm{eff} = 54.2 \pm 8.3$ and $63.6 \pm 9.4$ for FPI~1 and FPI~2, 
respectively.

%-------------------------------------------------------------------------------
%    Figure 
%-------------------------------------------------------------------------------
%-------------------------------------------------------------------------------

%-------------------------------------------------------------------------------
%    Table 3 
%-------------------------------------------------------------------------------
\begin{SCtable*}
\caption{Mean value and standard deviation of the effective finesse
    $F_\mathrm{eff}$ for FPI~1 and FPI~2 computed for three beam settings. 
    The standard deviation refers to the variation across the etalons rather
    than to a measurement error.}
\footnotesize   
\begin{tabular}{lcc}
\hline\hline
   & FPI~1 & FPI~2 \rule[-6pt]{0pt}{18pt}\\
\hline 
40~mm        & $51.8 \pm 7.4$ & $63.2 \pm 9.2$\rule[0pt]{0pt}{12pt}\\
65~mm        & $53.8 \pm 9.2$ & $63.7 \pm 9.5$\\
GREGOR pupil & $54.2 \pm 8.3$ & $63.6 \pm 9.4$\rule[-5pt]{0pt}{10pt}\\
\hline
\end{tabular}
\label{TAB03}
\end{SCtable*}
%-------------------------------------------------------------------------------

\begin{figure}[ht]
\centering
\includegraphics[width=0.95\textwidth]{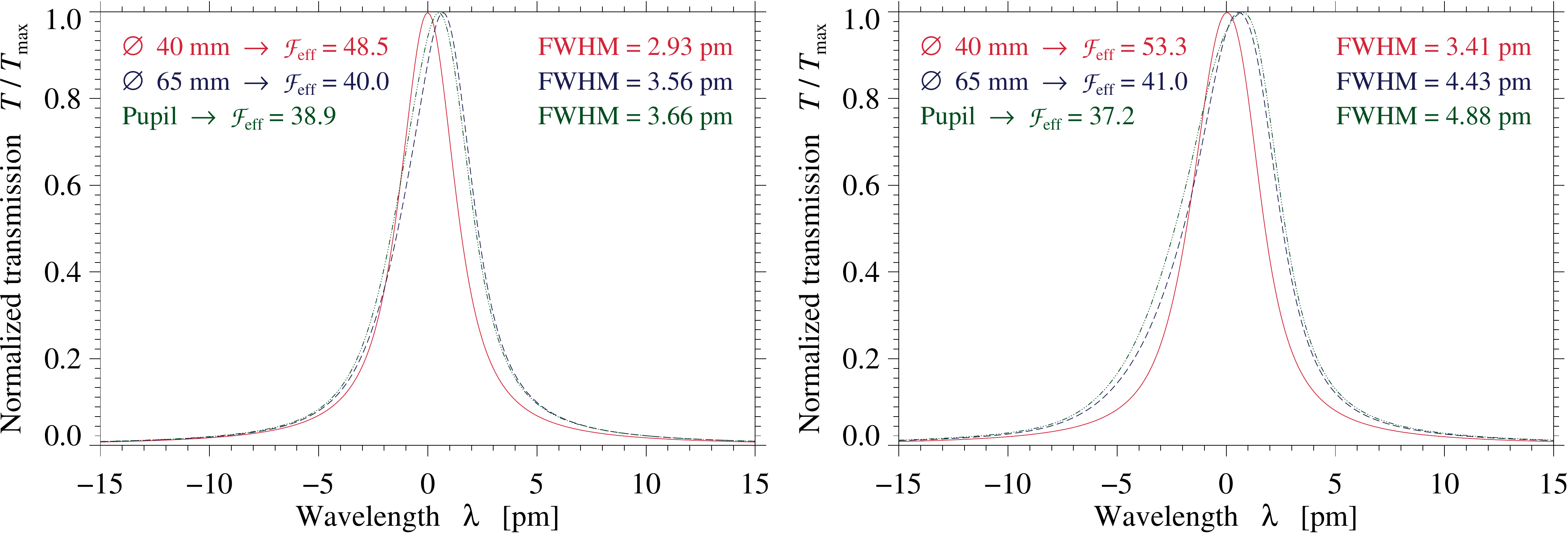}
\caption{Simulated transmission profiles of FPI~1 (\textit{left}) and 
    FPI~2 (\textit{right}) modeling the laser alignment procedure for the
    GFPI etalons. The profiles are shown for three beam settings: 40~mm 
    (\textit{solid red}), 63~mm GREGOR pupil (\textit{dash-dotted green}), and 
    65~mm (\textit{dashed blue}).}
\label{FIG09}
\end{figure}

The measured FWHM or finesse across the etalons facilitates simulating the 
transmission profiles of FPI~1 and FPI~2 as recorded in the GFPI alignment 
procedure described in Section~\ref{GFPI}, which also yields the effective 
finesse ${\cal F}_\mathrm{eff}$ for both etalons. In addition, the reflectivity 
of both etalons ($R_1 = 95.91$\% and $R_2 = 94.99$\%) is needed at the laser 
wavelength $\lambda$632.8~nm, which is given in the coating curves provided by 
the manufacturer. In the simulation, a clean Gaussian beam as described in 
Section~\ref{EVAL}, i.e., no interference fringes and no artifacts from in- and 
out-of-focus dust particles, provides the illumination of the etalons. The peak 
transmission, blue-shift, and FWHM across the etalons are taken from the 
laboratory measurements and the Lorentzian fits to the local transmission 
profiles. The integrated transmission profile, as it will be recorded with the 
photo-multiplier, can be derived by keeping the transmission profile of one 
etalon fixed while the other etalon scans the modulated intensity distribution 
(varying blue-shift and peak transmission). The results for the unobscured 40-mm 
and 65-mm beams and the 63-mm beam with an imprint of the GREGOR pupil are 
summarized in Figure~\ref{FIG09}. Sampling just the central part of the etalons 
leads to a much higher effective finesse ${\cal F}_\mathrm{eff}$ because the 
local transmission profiles are averaged over a much narrower blue-shift range. 
In principle, the FWHM will also be narrower in the central part and thus 
increases the effective finesse. However, as demonstrated in Figure~\ref{FIG07}, 
the GFPI etalons show as somewhat atypical finesse distribution, which was 
attributed to the aging of the coatings. Therefore, the reduced average 
blue-shift in the central part of the etalons plays the dominant role in shaping 
the transmission profiles. The transmission profiles are no longer symmetric 
because illumination as well as blue-shift, peak transmission, and FWHM of both 
etalons introduce imperfections in real-world applications. Finally, the 
observed transmission profiles can be approximated by Voigt profiles, where the 
Lorentzian part is a good approximation of the Airy function and the Gaussian 
part can be related to the microroughness errors \cite{Reardon2008}. The 
wavelength shifts are characterized by the standard deviation of the Gaussian, 
which are directly proportional to spacing fluctuations, i.e., the 
microroughness cavity error are $\sigma^m = 0.94$~nm and 0.78~nm for FPI~1 and 
FPI~2, respectively. These values are slightly higher but comparable to other 
studies \cite{Reardon2008}.

%-------------------------------------------------------------------------------
%    Gaussian Laser Beam
%    /home/denker/finesse/pro/finesse_beam20180725.pro
%-------------------------------------------------------------------------------
%-------------------------------------------------------------------------------

The collimator lenses for the GFPI laser channel ($f = 310$~mm) and the 
experimental setup at the VTT optical laboratory ($f = 800$~mm) introduce a 
different blue-shift. Thus, the simulated transmission profiles for the laser 
alignment procedure will differ in both setups and are not directly comparable. 
However, the significant broadening of the transmission profiles, which is 
introduced by averaging all blue-shifted profiles across the beam aperture is 
clearly apparent. Thus, the effective finesse ${\cal F}_\mathrm{eff}$ derived 
in the GFPI alignment procedure underestimates the effective finesse 
$F_\mathrm{eff}$ based only on the average of the local FWHM measurements. This 
difference was not always clearly expressed in literature.

%-------------------------------------------------------------------------------
%    Discussions
%-------------------------------------------------------------------------------

\section{Discussion}
\label{DISC}

In this work, we performed a quality assessment of the GFPI etalons. These 
etalons were in use for more than 10 years. Both etalons and their coatings are 
still in a very good condition showing only a minor degradation of the coatings. 
The average finesse values exceed those provided by manufacturer, even when 
taking the aging of etalon coatings into account. However, the finesse across 
the etalons exhibits an atypical radial profile with higher values at the edges 
rather than in the center as observed for other Fabry-P\'erot etalons. In 
addition, FPI~2 is the first etalon in the optical path encountering direct 
sunlight, and an imprint of the rotating GREGOR pupil is present and clearly 
evident in the two-dimensional maps of finesse, peak transmission, and FWHM. We 
attribute this to photo-chemical reactions in the coatings when exposed to 
sunlight and UV radiation. The order-sorting interference filter are 
typically blocked at the $10^{-4}$ level for UV radiation. However, the UV 
exposure over more than 10 years of observations will accumulate, and some 
alignment procedures even require direct exposure of the etalons to sunlight.
In addition, discrepancies of the average effective finesse are observed, 
especially for FPI~1, when compared with values derived from coating curves 
provide more than 10 years ago by the manufacturer. Our approach to the  
characteristic evaluation of Fabry-P\'erot etalons is not new 
\cite{Denker2005a}. However, the Zernike decomposition of the two-dimensional 
maps of characteristics is a powerful tool visualizing optical aberrations in 
imaging spectrometers, which potentially affect the instrument profile. This 
provided the motivation to carry out the long overdue characteristic evaluation 
of the GFPI etalons. The present findings are in good agreement with previously 
published results for other instruments \cite{Denker2005a, Reardon2008}. 
However, the imprint of the GREGOR pupil necessitated fitting high-order Zernike 
polynomials to the finesse of FPI~2, i.e., up to mode-ordering number $j=55$.

% Discussion about different beam size to measure finesse.
Another issue discussed in the present work is related to the operation and 
calibration of the GFPI, i.e., the impact of various beam sizes and shapes on 
measuring the plate parallelism. The limited space inside the the GFPI's 
protective housing and on top the optical tables restricted the diameter of the 
collimated laser beam to only 40~mm, which may not be sufficient to represent 
the overall tilt of the plates in the measuring procedure. Thus, three different 
configuration were evaluated in the experiments that were carried out in the 
optical laboratory of the VTT: (1) the current setup of the GFPI with a beam 
diameter of 40~mm, (2) a much large beam with a diameter of 65~mm, and (3) an 
arrangement that mimics the GREGOR pupil with an outer diameter of 63~mm. The 
various etalons properties are listed in Tables~\ref{TAB02} and~\ref{TAB03} for 
all three beam configurations. The differences are rather small. In principle, 
using only inner 40~mm of the etalons to determined the overall tilt of the 
plates is sufficient and justifies the current calibration procedure of the 
GFPI.

%Discussions about ghosts due to inter-reflections of two FPIs 
Inter-etalon reflections in Fabry-P\'erot instruments can create ghosts, which 
are usually eliminated by tilting the etalons.  However, tilting the etalons can 
introduce secondary affects \cite{Cavallini2006}. In the optical alignment of 
the GFPI ghosts are eliminated by ensuring that the surface of FPI~2 is 
perfectly perpendicular to the optical axis of the system, i.e., the 
retro-reflection of a pinhole in the intermediate focus F3 of the GREGOR 
telescope has to centrally pass an iris stop in the science focus F4, which is 
one of the fiducial points for the optical alignment of the GFPI. The other 
fiducial point is the first pupil image within the GFPI transfer optics. The 
second step is to tilt FPI~1 until all ghost images, e.g., those of a pinhole, 
are precisely stacked on top of each other. These calibration steps were omitted 
when individually evaluating the etalons in optical laboratory of the VTT 
because of the simplicity of the optical setup. However, eliminating ghosts is a 
critical step in the proper alignment of any Fabry-P\'erot instrument.

All etalons suffer from cavity errors, which have to be analyzed in detail 
\cite{Reardon2008}. Cavity errors can be separated in two classes, i.e., 
large-scale and randomly distributed errors. The random errors are caused by 
coating defects or the roughness of the plate surfaces. The large-scale plate 
defects are related to errors in figuring the plates, deformations caused by 
stresses in the coatings, and distortions introduced by the plate mountings. The 
latter cavity errors can be estimated and described using the Zernike polynomial 
fitting method \cite{Denker2005b}. The observed high mode number $j = 55$ 
suggests distortions of the coatings of the FPI~2 plates, which in turn are 
likely related to aging of these thin dielectric coatings when exposed to 
sunlight (especially UV radiation) for long periods of time. More recently, 
the Zernike polynomial fitting technique was adapted and expanded to accurately 
measure the cavity defects and more importantly to maintain the shape of the 
cavity while performing spectral scans \cite{Greco2019}. This new method 
incorporates concepts from phase-shifting interferometry and can be used for the 
any FPI that is tuned by changing the cavity spacing. The present assessment 
results of the GFPI etalons are encouraging in a sense that the overall quality 
of etalons remains high after more than a decade. However, etalons and their 
coatings should be regularly inspected every few years when used in 
campaign-style observations and even more frequently when operated on a daily 
basis, as expected for Fabry-P\'erot instruments of the next generation of 
4-meter aperture telescopes \cite{Tritschler2016, Matthews2016, Jurcak2019}.

%-------------------------------------------------------------------------------
%        Disclosure
%-------------------------------------------------------------------------------

\subsection*{Disclosures}

The authors declare that they have no conflicts of interest.

%-------------------------------------------------------------------------------
%        Acknowledgments
%-------------------------------------------------------------------------------

\acknowledgments The 1.5-meter \textit{GREGOR} solar telescope was built by a 
German consortium under the leadership of the Kiepenheuer Institute for Solar 
Physics in Freiburg with the Leibniz Institute for Astrophysics Potsdam, the 
Institute for Astrophysics G\"ottingen, and the Max Planck Institute for Solar 
System Research in G\"ottingen as partners, and with contributions by the 
Instituto de Astrof\'{\i}sica de Canarias and the Astronomical Institute of the 
Academy of Sciences of the Czech Republic. This research has made use of NASA's 
Astrophysics Data System. This study was supported by grant DE~787/5-1 of the 
Deutsche Forschungsgemeinschaft (DFG) and by the European Commission’s Horizon 
2020 Program under grant agreements 824064 (ESCAPE -- European Science Cluster 
of Astronomy \& Particle Physics ESFRI Research Infrastructures) and 824135 
(SOLARNET -- Integrating High Resolution Solar Physics). We would like to 
thank the referees who provided helpful comments improving the manuscript.

%-------------------------------------------------------------------------------
%        Bibliography
%-------------------------------------------------------------------------------

\listoffigures

%-------------------------------------------------------------------------------
%        Authors Biography
%-------------------------------------------------------------------------------

\vspace{2ex}\noindent\textbf{Meetu Verma} is a post-doctoral researcher at the 
Leibniz Institute for Astrophysics Potsdam (AIP). She received her doctoral 
degree in physics from the University Potsdam in 2013. Thereafter, she worked 
as a post-doctoral researcher at Max Planck Institute for Solar System Research in 
G\"ottingen and returned one year later to AIP. Her current research interests 
cover high-resolution observations of solar magnetic features, in particular, 
multi-instruments and multi-wavelength studies of sunspots in various stages of 
evolution. She organized and participated in many observing campaigns, in 
particular coordinated campaigns including space missions and German/U.S. solar 
telescopes.

\vspace{2ex}\noindent\textbf{Carsten Denker} is head of the ``Solar 
Physics" section and the solar observatory ``Einstein Tower" at the 
Leibniz Institute for Astrophysics Potsdam (AIP). He holds a doctoral degree in 
physics and a diploma in social sciences from the Georg-August University 
G\"ottingen. He regularly teaches astronomy and astrophysics lectures as an 
adjunct professor at the University Potsdam with an emphasis on solar and 
stellar physics. He is engaged in research and development of instruments for 
high-resolution solar physics and serves as the instrument PI of GFPI and HiFI. 
His research efforts concentrate on photospheric and chromospheric magnetic 
fields and activity, fine structure of sunspots, space weather, imaging 
spectropolarimetry, and image restoration.

%\listoftables

\end{spacing}

\end{document}